# Universal Numbers Library: design and implementation of a high-performance reproducible number systems library

Theodore Omtzigt
*Principal*
*Stillwater Supercomputing, Inc.*
El Dorado Hills, CA
theo@stillwater-sc.com

Peter Gottschling
*Principal*
*Simunova UG*
Leipzig, Germany
peter.gottschling@simunova.com

Mark Seligman
*Principal*
*InPredict*
Seattle, WA
stochasticboy@gmail.com

Bill Zorn
*Department of Computer Science*
*University of Washington*
Seattle, WA
billzorn@cs.washington.edu

*Abstract*— With the proliferation of embedded systems requiring intelligent behavior, custom number systems to optimize performance per Watt of the entire system become essential components for successful commercial products. We present the *Universal Number Library*, a high-performance number systems library that includes arbitrary integer, decimal, fixed-point, floating point, and introduces two tapered floating-point types, *posit and valid* that support reproducible arithmetic computation in arbitrary concurrency environments. We discuss the design of the Universal library as a run-time for application development, and as a platform for application-driven hardware validation. The library implementation is described, and examples are provided to show educational examples to elucidate the number system properties, and how specialization is used to yield very high-performance emulation on existing x86, ARM, and POWER processors. We will highlight the integration of the library in larger application environments in computational science and engineering to enable multi-precision and adaptive precision algorithms to improve performance and efficiency of large scale and real-time applications. We will demonstrate the integration of the Universal library into a high-performance reproducible linear algebra run-time. We will conclude with the roadmap of additional functionality of the library as we are targeting new application domains, such as Software Defined Radio, instrumentation, sensor fusion, and model-predictive control. (*Abstract*)

*Keywords—floating point, posit, reproducible linear algebra, C++, software, number system (key words)*

## I. INTRODUCTION

Mobile and cloud computing both are commercial endeavors at scale. In these environments, product performance is paramount as a 10% improvement in application performance or device power consumption can yield hundreds of millions of dollars in profit. Traditional High-Performance Computing (HPC) blazed the trail for efficient algorithms, but never had commercial scale, so power efficiency was never a design requirement. Yet, mobile and cloud computing do need those power efficiency improvements, and over the past decade we have seen vendors like Google and Microsoft ditch IEEE Floating Point for their next generation workloads, even building custom hardware to further their competitive advantage over their competitors. These next generation workloads also exhibit tremendous parallelism, and IEEE Floating Point rounding rules cause the loss of the associative and distributive laws of algebra. Algorithms need to be rewritten to deliver reproducibility. A new tapered floating-point number system, called *posit*, was specifically developed to solve this problem and at the same time improve on the information density of the encodings of the Reals, which will help improve power efficiency.

The proliferation of high-performance computing into real-time and embedded use cases has amplified this major short coming of the standard floating-point number system. High-performance task-level parallel systems introduce different execution orders of the original equations causing non-deterministic reordering of intermediate results. When such systems are inspected the non-deterministic reordering makes reproduction of the failure difficult if not impossible.

Several solutions to this problem have been proposed starting in 1986 by the work of Ulrich Kulisch[4]. The basic idea in that approach is to leverage a super-accumulator that accumulates intermediate results of a computational path at full precision. The actual rounding decision is made explicit by language constructs under control of the programmer. The fundamental problem in that approach is caused by the structure of floating point: a fixed-point representation of the result of a floating point multiply requires $1 + 2 \times (2^{e_{bits}} + m_{bits})$, where $e_{bits}$ and $m_{bits}$ are the number of bits in the exponent and mantissa, respectively. To be able to accumulate $2^k$ products we would add $k$ bits to the accumulator. For single precision and double precision floats the approximate size of these super-accumulators would be 640bits, and 4288 bits, respectively. Modern implementations of the Kulisch accumulator idea can be found in[5].

Another approach is to use arbitrary precision arithmetic. An example is the GNU Multiple Precision Floating-Point Reliably (MPFR [6]). The upside is that difficult computation problems in computational geometry and optimization become feasible, but the downside is that the common case is slowed down by 3 orders of magnitude.

ExBLAS [8][8] is a software approach that is not as slow as arbitrary precision but is still at least an order of magnitude slower than native execution. ExBLAS uses the super-accumulator approach coupled with a clever trick to compute the rounding error of each operation. By keeping track of how error accumulates in the basic linear algebra subroutines they are able to create reproducible results.

RepoBLAS [9] instead focuses on performance and relaxes the exactness constraint to deliver reproducible results in task-parallel execution environments.

Floating-point based arithmetic error control is complicated by the structure of the number systems. Super-accumulators grow exceptionally large due to the disproportional dynamic range of floats compared to their precision. Gustafson [1] has been working on tapered number systems to regain control over efficient and productive error control. He coined the term universal numbers, or unums for short. Unums come in several types, the Type III unums are called posits [3] and are the basis of our adaptive tensor processing architecture.

## II. POSITS

John Gustafson provides a Mathematica notebook that implements a reference for posits [2], we learn the definition of universal numbers, or unums, for short: "Unums are for expressing real numbers and ranges of real numbers." There are two modes of operation, selectable by the user: *posit mode* and *valid mode*.

In *posit mode*, a unum behaves much like a floating-point number of fixed size, rounding to the nearest expressible value if the result of a calculation is not expressible exactly; however, the posit representation offers more accuracy and a larger dynamic range than floats with the same number of bits. We can refer to these simply as *posits* for short, just as IEEE 754 Standard floating-point numbers are referred to as *floats*.

In *valid mode*, a unum represents a range of real numbers and can be used to rigorously bound answers much like interval arithmetic does, but with several improvements over traditional interval arithmetic. In this paper, we will focus on posit arithmetic exclusively.

A posit is made up of four components: sign, regime, exponent, and fraction. A posit is specified by its size in bits, *nbits*, and the maximum number of exponent bits, *es*.

Suppose we view the bit string for a posit as a 2's complement signed integer, ranging from $-2^{n-1}$ to $2^{n-1} - 1$. Let $k$ be the integer presented by the regime bits, and $e$ the unsigned integer represented by the exponent bits, if any. If the set of fraction bits is $\{f_1, f_2, \cdots, f_{fs}\}$ possibly the empty set, let $f$ be the value represented by $1.f_1f_2\cdots f_{fs}$. Then the value of a posit is defined by the following equation:

$$x = \begin{cases} 0, p = 0, \\ \pm\infty, p = -2^{n-1} \\ sign(p) \times useed^k \times 2^e \times f, all\ other\ p. \end{cases}$$

The following figure shows these fields for a 16-bit posit with 3 exponent bits, referred to as posit<16,3>.

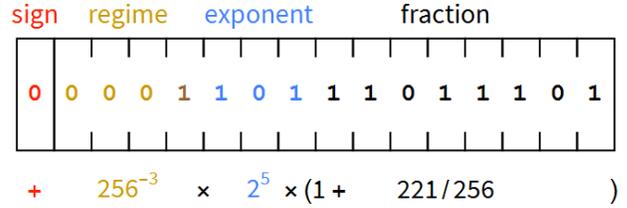

The sign bit 0, shown in red, implies that the value is positive. The regime bits have a run of three 0s terminated by the opposite bit 1, which implies the power of *useed* is -3. *Useed* is defined as $2^{2^{es}}$ and represents the scaling factor of the regime. In this example, the scale factor contributed by the regime is $256^{-3}$. The exponent bits 101, shown in blue, represent 5 as an unsigned binary integer, and contribute a scale factor of $2^5$. Finally, the fraction bit 11011101, shown in black, represent 221 as an unsigned binary integer, yielding a fraction value of $1.0 + 221/256$. The value of this posit bit pattern is $477 \times 2^{-27} \sim 3.55393 \times 10^{-6}$.

The size of the regime and exponent fields is variable creating a tapered precision real number system, with a dynamic range perfectly symmetric around 1. The minimum and maximum positive number for a posit configuration are called *minpos* and *maxpos*. Their values are a function of the scaling factor of the regime and the size of the posit:

$$\{minpos, maxpos\} = \{useed^{-nbits+2}, useed^{nbits-2}\}$$

The ratio of *maxpos* to *minpos* is $useed^{2nbits-4}$, which defines the dynamic range of the posit. The posit format uses regime bits to raise *useed* to the power of any integer from $-nbits+1$ to $nbits-1$, otherwise stated, the dynamic range of a posit is an exponential of an exponential of an exponential. This allows posits to create a larger dynamic range from fewer exponent bits than IEEE floats, leaving more fraction bits available to improve the precision of a value representation.

A wonderful way to visualize the structure of a posit configuration is to realize that they derive from Type II unums that mapped binary integers to the projective reals. Projective reals wrap the real number line onto a circle so that negative and positive infinity meet at the top.

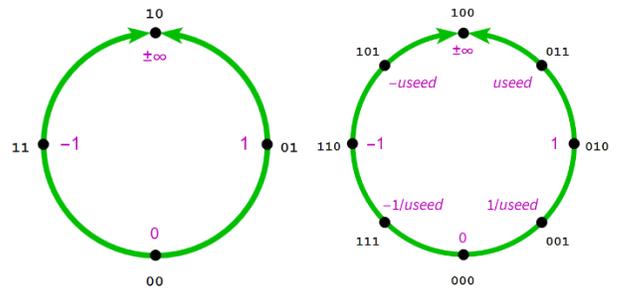

The diagram on the left represents a 2 bit posit. We move to three bits by inserting a value between 1 and $\pm\infty$. It could be any real number greater than 1; it could be *2, 10, $\pi$*, or *googol*. The choice of this number *seeds* how the rest of the ring of unums is populated, to signify its importance this number was given the symbolic name *useed*. As we have seen above, for posits this value is set to $2^{2^{es}}$. Further bit expansion follows the rules that negation reflects about the vertical axis, and reciprocation reflects about the horizontal axis. The next figure shows a ring plot of values for a posit<5,1>:

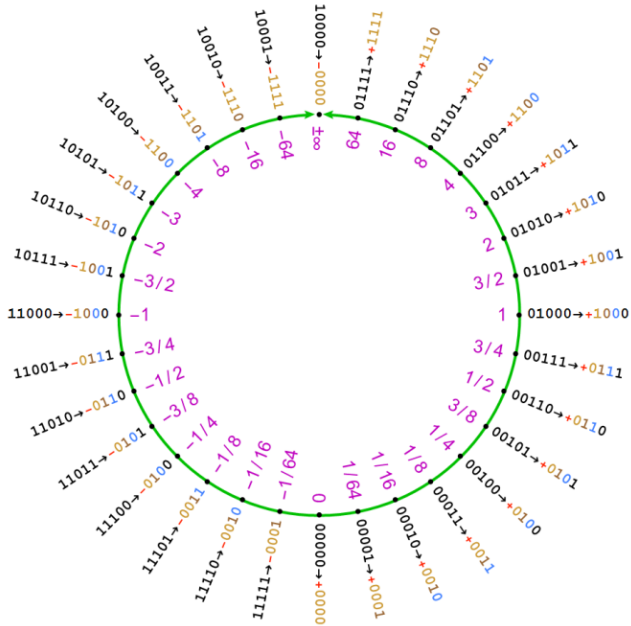

### III. A MODERN C++ LIBRARY FOR NUMBER SYSTEMS

The Universal library began with the following ease of use requirements:

- Modern C++ implementation
- No external dependencies, header-only template library
- Natural extension of native types: plug-in replacement of IEEE floats/doubles/long doubles
- Arbitrary configuration posits to support multi-precision algorithm design and experimentation
- Should be able to function as a hardware verification reference for custom hardware and soft-core implementations in ASICs and FPGAs.
- Available on Windows, MacOS, and Linux, and compatible with compilers that support C++14, MSVC, gcc, and clang. Intel C++, and executable on x86, ARM, and POWER processors.

The requirement to support arbitrary configuration posits without a validation reference implied that we needed to add that capability to the design. Posit configurations are defined by the size of the posit, defined by *nbits*, and the size of the exponent field, defined by *es*. Due to the nature of the posit encoding, both configuration parameters are independent. That is, you could define an 8-bit posit with a 10-bit exponent field. That implies that the full configuration space for posits is very large, and thus any validation suite would need to be parameterized as well.

We settled on implementing the first incarnation on a bit-level abstraction of the posit and its arithmetic. With that abstraction, all arithmetic can be defined in terms of *nbits,* and *es*. But more importantly, we can leverage induction to prove the validity of posit<*nbits+1,es*>, and thus the validation suite can match any posit configuration. By using fully enumerated state space validation for small configuration, such as posit<8,0>, we can use induction to validate that posit<9,0> is valid, and transitively, that the implementation of posit<256,5> is valid. By and large this has been confirmed, but we still encountered subtle bugs particularly in the rounding code to convert to native language types when the posit configuration has values that are not representable by IEEE floating point.

A second area of complexity was in generating the golden values for the validation reference. For example, for the standard 32-bit posit, posit<32,2>, we cannot use IEEE doubles to generate a validation reference, because IEEE doubles do not have enough precision bits to represent a posit<32,2> product. Further complicating this problem, the Windows platform does not support a long double format, and on the Linux platforms, long doubles are just extended precision doubles, so will also fail. We needed a reference floating point implementation that supported arbitrary precision floating point. We initially leveraged MPFR, but since that violates the no-dependencies requirement, we ended up implementing an arbitrary floating-point class within Universal as well.

The need to have an arbitrary linear floating-point library implementation was a first indication that the Universal library's narrow starting point to create an arbitrary posit configuration number system library had to expand. The internal posit arithmetic is implemented as a standard triple floating-point format of <sign, exponent, mantissa> with the conversion routines marshalling posit encodings to and from the triple format which the arithmetic functions use. But the validation suite requirements forced this implementation to have its own public interface.

The next event in this progression of expanding the Universal library to become a truly universal number system library came from the numerical experiments to quantify the benefits of the posit number system, and in particular, the benefits of reproducibility due to user-defined rounding offered by the quire. We selected Hilbert matrices as our test subject due to their well-known high condition numbers. One problem with these matrices is that in their original formulation, they contain coefficients that are not representable by a binary number system. To circumvent injecting rounding errors in the basic formulation, we scale the Hilbert matrix coefficients so that all coefficients can be represented by a binary number system. These scaling numbers grow very rapidly, thus creating the problem of having to represent exceptionally large integers, which were also needed to generate the algebraic Hilbert matrix inverses.

The latest requirement has been for arbitrary fixed-point number systems, from telecommunication and satellite communication and sensor networks. These systems tend to be implemented as fixed-point hardware, but more advanced signal processing algorithms are hitting constraints of dynamic range of fixed-point numbers. Posit numbers offer custom dynamic ranges at very small bit widths, and thus posits are a very powerful replacement for fixed-point pipelines when the signal processing complexity can be more easily supported.

As a consequence, the Universal library now supports arbitrary integers, decimals, fixed-point, floating point, posits and valids, and we have had requests for other number systems, such as log-based systems for Deep Learning, other tapered floating-point systems for HPC, block compressed numbers, and even custom number systems for quantum computing. Adding additional number systems is now a well understood pattern, and we invite collaborators to add their inventions to the library.

## IV. MATH LIBRARY

When creating a new arithmetic system, the library will also need to provide a custom math library for that arithmetic. This is a major undertaking with complex validation requirements. For posits, the goal is to have always correctly rounded results for all the functions in the math library. For specific posit configurations there are implementations for the core logarithm, exponential, and trigonometry functions but they use custom minmax functions. There is no solution yet to automatically parameterize these functions to support arbitrary posit configurations.

For application integrations, we have a fallback to use the standard math library using conversions to and from posit and long double. In practice it works but is clearly limited. It precludes us from leveraging the 128-bit and 256-bit posits. Providing a universal math library that adheres to the goal of posits to provide correctly rounded results will be the most active research area for the library.

## V. SPECIALIZATIONS FOR PERFORMANCE

One of the disadvantages of our reference implementation is that the additional detail to bit-level operators slows down the speed of arithmetic operators. Systems such as Softfloat and MPFR use integer operators to yield much higher performance. The Universal library provides integer operation implementations of a select set of posit configurations that are of interest to application domains.

C++ templates allow for template specializations and we leverage this for implementing fast integer algorithms for the standard posits, and some key posits that have shown value in Computer Aided Engineering and Deep Learning applications.

The performance of these fast implementations is about 100x slower than hardware IEEE floating-point but enables full scale application tests since the memory footprint of these posits is native. Table 1 shows the current performance of these fast posit implementations.

| Configuration | Performance Intel Xeon E5 2.2GHz Single Core | Notes |
|---|---|---|
| posit<4,0> | 1 GPOPS | 8-bit integer |
| posit<8,0> | 135 MPOPS | 8-bit integer math |
| posit<16,1> | 115 MPOPS | 16-bit integer |
| posit<32,2> | 110 MPOPS | 32-bit integer |
| posit<64,3> | 105 MPOPS | 64-bit integer |
| posit<128,4> | 1 MPOPS | Bit-level |
| posit<256,5> | 1 MPOPS | Bit-level |

*Table 1: software emulation performance of the standard posit configurations*

With the performance level of the Universal library, it is feasible to run large MPI applications to study the impact of accuracy and reproducibility when using posit arithmetic.

## VI. INTERFACES TO C AND FORTRAN ENVIRONMENTS

In addition to the C++ template library, Universal also provides a pure C library like Softfloat[10] and Softposit[11] for better compatibility with projects that do not use C++. Ideally, the C and C++ libraries would share the same core code for number system conversions and arithmetic; however, this is difficult to implement in practice.

One seemingly promising approach is to build the C API as a shim for the C++ template library. While this is a convenient way to access the C++ templates from within C code, it is not pure C, as the compiled templates still contain dependencies on the C++ standard library. Another approach is to build the core code in C, which can then be included in the C++ templates. In practice, this leads to an observed slowdown of about 15%, possibly due to less efficient in-lining by the compiler.

Maintaining separate, high-performance, pure C and C++ is challenging, particularly for datatypes like posits that can support a huge number of configurations. At present, Universal supports a subset of posits for its C API using an independent codebase. Going forward, a better approach would be to automatically generate C stubs to support custom types and operations.

## VII. APPLICATION RUN-TIME

The Matrix Tensor Library (formerly Matrix Template Library) [12][13] is a generic C++ library for sparse and dense linear algebra. It is designed to realize scientific algorithms as naturally as possible in C++ by establishing a domain-specific embedded language (DSEL) for an intuitive notation. At the same time, it provides high performance due to advanced programming techniques like expression templates and by interfacing existing high-performance libraries.

The currently developed version 5 (MTL5) integrates posits without the need of further user interaction. The generic design allows for creating vectors, matrices, and other tensors with posit values. For instance, a posit matrix can be initialized with a file from the matrix market – by simply passing the file name to the matrix constructor. All MTL5 operations work out of the box with posit numbers.

Thus far, two operations are specialized for posits:
- dot product and
- product of a row-major matrix with a vector.

In those algorithms, the quire is used in all intermediate results and only the result is rounded. For instance, the computation of:

$$\left\langle \begin{pmatrix} 1 \\ 1^{-20} \\ -1 \end{pmatrix}, \begin{pmatrix} 1 \\ 1 \\ 1 \end{pmatrix} \right\rangle$$

yields the value 0 when using a 64-bit IEEE (i.e., C++ type double), but the correct value with posit<32,2>; the nearest representable value of $9.99896 \cdot 10^{-21}$.

The benefit of posits and the quire compared to IEEE-754 arithmetic manifested clearly when solving linear systems iteratively. As iterative solver we used a Conjugate Gradient method with an ILU preconditioner. The matrices originated from the Harwell-Boeing-Collection in the matrix market. The right-hand-side vector *b* was set to $A \cdot \mathbf{1}$ and the initial value of *x* to **0**. The choice of b allowed us to measure the reduction of the error (instead of merely measuring the reduction of the residual).

With this scenario we observe:
- Faster convergence: With matrix bcsstk14.mtx and a requested error reduction of $10^{-11}$, the solution was found after 397 iterations with double values compared to 288 iterations with posit<64,3>.
- Higher error reduction: For solving a linear system with matrix bcsstk16.mtx, error was reduced by a factor of $6.10129 \cdot 10^{-16}$ after 61 iterations with posit<64,3> whereby double only allowed for an error reduction of $3.3 \cdot 10^{-15}$ (after 62 iterations).
- Preserving the result: When requesting an error reduction of $10^{-13}$ with bcsstk16.mtx, the system was solved after 56 iterations with double or posit<64,3>. With float values the error was reduced by $3.115 \cdot 10^{-6}$ after 44 iterations. The error remained similar for some time and turned into –NaN after 125 iterations. The requested precision was not achievable with posit<32,2> either, but the error could be reduced by $7.488 \cdot 10^{-7}$ and subsequent computations preserved this result.

These early results demonstrate the benefits of using posits and deferred rounding techniques to improve the performance of complex linear algebra operations. This will be an active research area for a while as we are exploring multi-precision and reproducible algorithms for iterative solvers, direct solvers, eigen value solvers, and convex and non-convex optimization.

## VIII. AUTOMATED CI/CD

As an open-source project, to enable high-productivity collaboration, we implemented a Continuous Integration and Continuous Deployment capability before the first external collaborators were invited. The CI system is built with CodeShip[14] and AppVeyor[15] and will compile and test on any code pull request (PR). The CI build and test is accomplished through a build/test container pattern. We have a Docker container that contains a build environment which gets run on each PR. The build container generates the test executables, which are then passed on to a test container which will run the regression suite. If either the build container or the test container fails, the PR will be flagged. The CodeShip build containers support gcc and clang compiler suites, and we added AppVeyor to the CI pipeline as it supports Windows containers and compiler environments.

The regression suite consists of roughly 160 validation tests ranging from arithmetic and logic tests to coding patterns and educational examples. Any change to the API or functionality will likely trigger a failure in this regression suite, thus protecting new developers of taking the arithmetic systems down due to inexperience: the CI system will flag these errors automatically. On a reasonably modern machine, the regression suite runs in about 10mins.

## IX. CONTAINER

The Universal library is available as a container. When you just want to explore the functionality without having to clone and build the environment, a Docker container is available. We release new functionality regularly, and the container contains some useful command line tools to look at bit patterns of IEEE floating point and posit conversions.

The docker container can be run with the following command:
> *docker run -it –rm stillwater/universal /bin/bash*

More information can be found at the Universal github repo [16].

## X. MULTI-PRECISION RESEARCH

Given the broad functionality of the number systems available in the Universal library, it is a very productive environment to research multi-precision algorithms. We mentioned this briefly before, Hilbert matrices are fun but particularly challenging test subject. A Hilbert matrix, introduced by David Hilbert (1894), is a square matrix with entries being the unit fractions:

$$H_{ij} = \frac{1}{i+j-1}$$

For example, this is the $5 \times 5$ Hilbert matrix:

$$\begin{bmatrix} 1 & \frac{1}{2} & \frac{1}{3} & \frac{1}{4} & \frac{1}{5} \\ \frac{1}{2} & \frac{1}{3} & \frac{1}{4} & \frac{1}{5} & \frac{1}{6} \\ \frac{1}{3} & \frac{1}{4} & \frac{1}{5} & \frac{1}{6} & \frac{1}{7} \\ \frac{1}{4} & \frac{1}{5} & \frac{1}{6} & \frac{1}{7} & \frac{1}{8} \\ \frac{1}{5} & \frac{1}{6} & \frac{1}{7} & \frac{1}{8} & \frac{1}{9} \end{bmatrix}$$

The Hilbert matrix has an algebraic inverse expressed using binomial coefficients:

$$H^{-1}{}_{ij} = (-1)^{i+j}(i+j-1)\binom{n+i-1}{n-j}\binom{n+j-1}{n-i}\binom{i+j-2}{i-1}^2$$

To calculate the Hilbert inverses, the binomial coefficients require significant dynamic range, but are easily computed with arbitrary size integers. We use the *integer<nbits>* class to generate a validation reference for our Hilbert matrices to make certain that we do not clip coefficients due to our number system lacking enough dynamic range.

A second example is the study of minimum precision number systems to capture the dynamics of chaotic systems. The Lorenz system is a system of ordinary differential equations, notable for having chaotic solutions. It was developed by Edward Lorenz as a simplified mathematical model for atmospheric convection [21] described by the following system of ordinary differential equations.

$$\frac{dx}{dt} = \sigma(y - x)$$
$$\frac{dy}{dt} = x(\rho - z) - y$$
$$\frac{dz}{dt} = xy - \beta z$$

The basic idea is that imprecision in the number system will generate different trajectories in state space of the evolution of this ODE. Chaotic systems, and in particular this Lorenz system, exhibit regions of state space that appear stable and predictable called attractors, and rapid transitions between attractors that are extremely sensitive to perturbations. These perturbations can come from rounding error in the underlying number system, and thus need to be avoided.

We can generate a reference trajectory using a precise number system, and then explore less precise number systems that will execute at much higher performance with the goal to still generate the same trajectory as the reference solution.

Figure 1 shows the output of a collection of empirical experiments using ODEintv2 [19] to solve the Lorenz ODE with a $4^{th}$ order Runge-Kutta solver and different posit configurations from the Universal library. Odeint is a modern C++ library for numerically solving Ordinary Differential Equations. It is developed in a generic way using Template Metaprogramming which leads to extraordinary high flexibility at top performance. The numerical algorithms are implemented independently of the underlying arithmetic.

The reference trajectory is provided by solving the Lorenz system with a high precision posit<64,3>. As one can clearly see, a posit<16,1> is not quite sufficient to capture the dynamics of this system, but a posit<18,1> can faithfully reproduce the reference solution.

## XI. APPLICATION INTEGRATIONS

There are several notable application integrations of the Universal library to offer posit arithmetic to study accuracy and reproducibility of advanced applications in computational science and engineering.

The first integration of posits in a large-scale application code was G+Smo [17] and lead by Matthias Moller at Delft University of Technology. G+Smo (Geometry + Simulation Modules, pronounced "gismo") is an open-source C++ library that brings together mathematical tools for geometric design and numerical simulation. It implements the relatively new paradigm of iso-geometric analysis, which suggests the use of a unified framework in the design and analysis pipeline. G+Smo is an object-oriented, cross-platform, template C++ library and follows the generic programming principle, with a focus on both efficiency and ease of use. The library has been developed within the homonym research network supported by the Austrian Science Fund and aims at providing access to high quality, open-source software to the forming iso-geometric numerical simulation community and beyond.

The second large scale integration was in TVM/VTA and developed by Gus Smith. TVM is an open deep learning compiler stack for CPUs, GPUs, and specialized accelerators. It aims to close the gap between the productivity-focused deep learning frameworks, and the performance- or efficiency-oriented hardware backends. TVM provides the following main features:

- Compilation of deep learning models in Keras, MXNet, PyTorch, Tensorflow, CoreML, DarkNet into minimum deployable modules on diverse hardware backends.
- Infrastructure to automatic generate and optimize tensor operators on more backend with better performance.

TVM stack began as a research project at the SAMPL group of Paul G. Allen School of Computer Science & Engineering, University of Washington. The project is now driven by an open-source community involving multiple industry and academic institutions.

The most recent integration is Autodiff [20] lead by Allan Leal. Autodiff is a C++17 library that uses modern and advanced programming techniques to enable automatic computation of derivatives in an efficient and easy to use way. It supports both forward and reverse modes. In a *forward mode automatic differentiation* algorithm, both output variables and one or more of their derivatives are computed together. For example, the function evaluation *f(x, y, z)* can be transformed in a way that it will not only produce the value of *u, the output variable*, but also one or more of its derivatives *(∂u/∂x, ∂u/∂y, ∂u/∂z)* with respect to the *input variables (x, y, z)*. In a *reverse mode automatic differentiation* algorithm, the output variable of a function is evaluated first. During this function evaluation, all mathematical operations between the input variables are *"recorded"* in an *expression tree*. By traversing this tree from top-level (output variable as the root node) to bottom-level (input variables as the leaf nodes), it is possible to compute the contribution of each branch on the derivatives of the output variable with respect to input variables.

## ACKNOWLEDGMENT

Special thanks go out to John Gustafson for creating the universal numbers and his tireless support for answering our questions and helping us with examples and detailed references.

Special thanks to Matthias Moller for the integration of posits into G+SMO, Gus Smith for TVM integration, Allan Leal for integration into Autodiff, Mario Mulansky for help with ODEintv2.

Finally, thanks to the open-source community for contributing bug fixes and filing issues and feature requests.

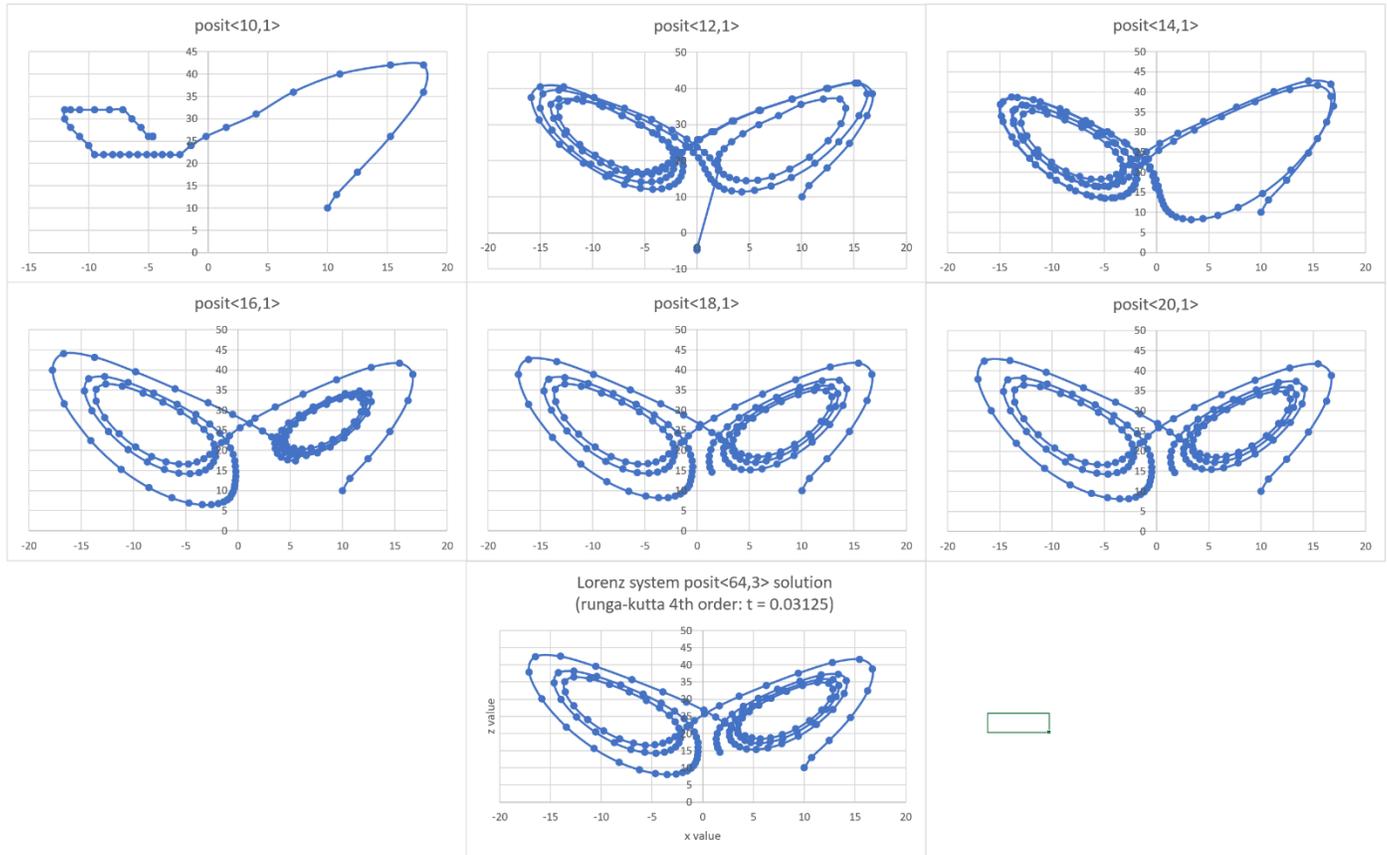

*Figure 1 Multi-precision algorithm experiments for solving the Lorenz System ODE*